# Analysis of Student Satisfaction Toward Quality of Service Facility

View the article online for updates and enhancements.

# Analysis of Student Satisfaction Toward Quality of Service Facility

D Napitupulu[1]*, R Rahim[2], D Abdullah[3], MI Setiawan[4], LA Abdillah[5], AS Ahmar[6], J Simarmata[7], R Hidayat[8], H Nurdiyanto[9], A Pranolo[10]

[1]Research Center for Quality System & Testing Technology,
Indonesian Institute of Sciences, Tangerang Selatan, Banten, Indonesia
[2]School of Computer and Communication Engineering, Universiti Malaysia Perlis, Malaysia
[3]Department of Informatics, Universitas Malikusaleh, Aceh, Indonesia
[4]Department of Civil Engineering, Narotama University, Surabaya, Indonesia
[5]Department of Information System, Universitas Bina Darma, Palembang, Indonesia
[6]Department of Statistics, Universitas Negeri Makassar, Makassar, Indonesia
[7]Department of Informatics, Universitas Malikusaleh, Aceh, Indonesia
[8]Department of Information Technology, Politeknik Negeri, Indonesia
[9]Department of Informatics Engineering, STMIK Dharma Wacana, Lampung, Indonesia
[10]Department of Informatics, Universitas Ahmad Dahlan, Indonesia

*darwan.na70@gmail.com

**Abstract.** The development of higher education is very rapid rise to the tight competition both public universities and private colleges. XYZ University realized to win the competition, required continuous quality improvement, including the quality of existing service facilities. Amenities quality services is believed to support the success of the learning activities and improve user satisfaction. This study aims to determine the extent to which the quality of the services effect on user satisfaction. The research method used is survey-based questionnaire that measure perception and expectation. The results showed a gap between perception and expectations of the respondents have a negative value for each item. This means XYZ service facility at the university is not currently meet the expectations of society members. Three service facility that has the lowest index is based on the perception of respondents is a laboratory (2.56), computer and multimedia (2.63) as well as wifi network (2.99). The magnitude of the correlation between satisfaction with the quality of service facilities is 0.725 which means a strong and positive relationship. The influence of the quality of service facilities to the satisfaction of the students is 0.525 meaning that the variable quality of the services facility can explain 52.5% of the variable satisfaction. The study provided recommendations for improvements to enhance the quality of services facility at the XYZ university facilities.

## 1. Introduction
These Quality education system of course closely related to the quality of human resources. It is based on the reality that the output of an educational system is the human resource that will be used in industry and development in an area. Efforts to improve the quality of human resources can be done by improving the educational system which can then produce a quality education system. The government is also aware of the importance of education, this can be seen by the implementation of 9-year compulsory education program for Indonesian citizens. Private universities are one of the alternatives for society to face the increasing demand of society to the education needs of today. Various advantages of private universities that will attract prospective students increasingly highlighted, ranging from degree bachelor, curriculum given to a variety of facilities that will be obtained by students will study on campus. Competition in the world of education is also getting tighter with the increase of universities, especially private universities (PTS) in Jakarta caused increasingly fierce competition. XYZ University is aware to win the competition, it needs continuous quality improvement including the quality of service facilities available to users. In addition to the continuously updated curriculum, service facilities also need to be improved for campus advancement [1].

The level of satisfaction is a function of the difference between perception and expectation [2]. The purpose of this study is to measure the gap between the perceptions and expectations of users of service facilities as a form of evaluation of the performance of the university. This research also wanted to know the relation between quality of service facility and user satisfaction. At the same time how much influence the quality of service facilities to user satisfaction. Practically, this research is expected to provide input for management related to the evaluation of condition of service facility based on user perspective so that it can formulated a comprehensive service quality improvement strategy in facing the tight competition in educational environment especially in Jakarta. Thus, in turn, it can attract high school / senior high school students to enroll in the campus environment.

Based on the background that has been described before, it can be formulated research problems is how the level of user satisfaction with the quality of existing service facilities? Then how the relationship between the quality of service facilities with user satisfaction on the XYZ university? Thus the purpose of this study is to evaluate the condition of service facilities based on user perspective so it can be known how far the level of user satisfaction of existing facilities. This can provide input for university management to make continuous improvement process.

## 2. Literature Review
If you Based on the literature review, similar research that has been done can be explained as follows:

- Research conducted by Sutino and Sumarno (2005) on the Effect of Product Quality (Service) and Quality of Service to Customer Satisfaction and Loyalty at PT. POS Indonesia Semarang. The results showed that the quality of service influence on satisfaction and satisfaction proved to affect loyalty [3].
- Research conducted by Samosir and Zurni (2005) on the Effect of Service Quality on Student Satisfaction Using USU Library. This research is descriptive explanatory where the method used is a survey method that uses questionnaires as a tool to collect data with the unit of analysis is USU students. The study used descriptive statistics and showed that service quality (reliability, responsiveness, assurance, empathy, and direct evidence) simultaneously had a significant effect on student satisfaction [4].
- Research conducted by Sutardji and Sri (2006) on the Analysis of Some Factors that Influence on Library User Satisfaction. The research method used is survey with questioneer as a tool to collect data. The results showed there was a positive influence between service quality variable to student satisfaction variable. Thus, the quality of service has a positive effect on student satisfaction [5].
- Research conducted by Sukandi (2010) on the Relationship Between Campus Facilities To Student Satisfaction In Facing the Competitiveness of Educational Services. This research is descriptive with survey method. The results showed that there is a positive relationship between the facilities provided by the campus on student satisfaction [6].

Based on previous literature review it can be said that there is a relationship between service quality and student satisfaction. Therefore, in this research want to know how big relation between service quality with student satisfaction and the influence of service quality to student satisfaction especially in XYZ university.

## 3. Research Methodology

In this section we The sample is part of the number and characteristics possessed by that population or the delivery of something by representatives of the class, but simply represents the whole opinion. This research was conducted by questionnaire-based survey method that was distributed to students of XYZ University Computer Science Faculty. Sampling technique with purposive sampling from all study program in Computer Science Faculty (S1), Information System (S1), Computer System (S1), Computer Engineering (D3) and Information Management (D3). Here is the distribution of active student data from the Faculty of Computer Science at XYZ University as the study population presented in table 1 below:

Table 1. Population of Respondent.

| No | Department | Number of Students |
|---|---|---|
| 1. | Computer Science | 213 |
| 2. | Information System | 425 |
| 3. | Computer System | 126 |
| 4. | Computer Engineering | 25 |
| 5. | Information Management | 53 |
|  | Total Population | 842 |

In this study the number of samples is calculated using the Slovin formula [7] where the population is the total active students academic year 2015/2016 is 842 people, namely:

$$\eta = \frac{N}{1 + (N \times e^2)} \quad (1)$$

where :
η = Number of samples
N = Population
e = Inaccuracy ease due to intolerable sampling error (using 10%)

Based on the formula (1) above, the sample number is as follows:

$$\eta = \frac{842}{1 + (842 \times 0.1^2)} = 89{,}38 \quad (2)$$

Thus the minimum sample size in this study was 89 respondents.

The distributed questionnaire has been designed and tested face validity by 30 respondents. Face validity is done to ensure the respondent understands the questions or statements contained in the questionnaire. Based on the results of face test (face validity) showed that all respondents agree that each item in the questionnaire easy to read and understand as in Table 2. The questionnaire distributed using Likert scale is the scale used to calculate a perception, attitude or opinion of a person or group about a Events or social phenomena based on operational variables that have been established by researchers. The scale used has intervals of 1-5 where 1 (very unsatisfied), 2 (unsatisfied), 3 (less satisfied), 4 (satisfied) and 5 (very satisfied). A total of 89 respondents who were involved only 84 respondents who answered and filled out the research questionnaire completely. An incomplete questionnaire will not be included in the analysis phase.

## 4. Result and Discussions

### 4.1. Validity and Reliability Testing

Before the data is processed further, the validity and reliability test of the questionnaire was conducted to determine the extent to which the measuring instrument is accurate and reliable or reliable. Testing the validity and reliability of the questionnaire was done with the help of the program SPPS IBM Statistics version 22 where obtained correlation coefficient (Corrected Item-Total Correlation) above 0.3 means that all items questionnaire can be said valid [8]. While the reliability coefficient is determined based on the value of Cronbach's Alpha where obtained the overall value is 0.83, which means that the questionnaire used has been reliable because it meets the minimum requirement of 0.60 [9]. From the results of validity and reliability test can be said that the questionnaire used in this study is accurate and reliable. After validity and reliability test, based on the data that has been processed, the analysis of user satisfaction level through measurement to the perception and expectation of the users on the quality of campus service facilities that exist as presented in Table 2.

### 4.2. User Satisfaction Analysis

In Table 2, there are two dimensions of learning space dimension and campus environment dimension. The dimension of the classroom is related to the facilities that support learning activities in the classroom. While the dimension of the campus environment is a facility outside the classroom that supports learning activities for students. Based on Table 3 can be seen gap between the perceptions and expectations of respondents on the quality of existing service facilities. From the result of measurement of satisfaction, it seems that for each item questionnaire, perception index is smaller than user expectation index. This means that the quality of service facilities that exist today is not yet satisfactory users. When further examined, the gap between perceptions and expectations is negative (perceptions - expectations) which means perceived performance is still far from expected.

**Table 2.** Perception and Expectation Measurement.

| No | Item | Perception | Expectation |
|---|---|---|---|
| | Class Room Dimension | | |
| 1. | Air Conditioning | 3.79 | 4.12 |
| 2. | Facilities benches and chairs in the classroom | 3.39 | 4.37 |
| 3. | Computer and multimedia facilities in the classroom | 2.63 | 4.25 |
| 4. | Stationary facilities in the classroom | 3.95 | 4.06 |
| 5. | Facilities of books - books, other reference journals | 3.01 | 3.84 |
| 6. | Laboratory facilities on campus | 2.56 | 4.30 |
| | Environment Dimension | | |
| 7 | Wifi network facilities on campus | 2.99 | 4.38 |
| 8. | Library room facilities on campus | 3.49 | 4.15 |
| 9. | Facility of Student Activity Unit on campus | 3.23 | 4.09 |
| 10. | Sports activities facilities on campus | 3.35 | 4.18 |
| 11. | Vehicle parking facilities on campus | 3.27 | 4.23 |
| 12. | College canteen facility | 3.24 | 4.21 |
| 13. | Convenience learning situations that can motivate student learning | 3.69 | 4.45 |
| 14. | Cleaning facilities Campus toilets | 3.24 | 4.13 |
| | Average score | 3.27 | 4.20 |

Table 3. User Satisfaction Level.

| No | Item | Gap |
|---|---|---|
| 1. | Air Conditioning | -0.33 |
| 2. | Facilities benches and chairs in the classroom | -0.98 |
| 3. | Computer and multimedia facilities in the classroom | -1.62 |
| 4. | Stationary facilities in the classroom | -0.11 |
| 5. | Facilities of books - books, other reference journals | -0.83 |
| 6. | Laboratory facilities on campus | -1.74 |
| 7. | Wifi network facilities on campus | -1.39 |
| 8. | Library room facilities on campus | -0.66 |
| 9 | Facility of Student Activity Unit on campus | -0.86 |
| 10. | Sports activities facilities on campus | -0.83 |
| 11. | Vehicle parking facilities on campus | -0.96 |
| 12. | College canteen facility | -0.97 |
| 13. | Convenience learning situations that can motivate student learning | -0.76 |
| 14. | Cleaning facilities Campus toilets | -0.89 |
| | **Average Score** | **-0.92** |

The total user satisfaction level also shows a negative value of -0.92. This indicates that overall users feel the condition of existing service facilities is still less or have not met the expectations of users. In fact, there are 3 (three) service facilities that have the lowest perception index of laboratory facilities (2.56), computer facilities and multimedia in the classroom (2.63) and wifi network facilities (2.99). Whereas the laboratory, computer and multimedia and internet network is a very important infrastructure in developing the core competencies of computer science faculty students.

*4.3. Correlation of User Satisfaction and Service Facility*

In this study wanted to know the relationship between the variable quality of campus service facilities with student satisfaction. The size of the relationship between variables is expressed in numbers called the correlation coefficient (R) that is between -1 to +1 where if approaching +1 then the relationship is stronger and positive. Meanwhile, if close to -1 then there is a stronger relationship but the direction is negative. If the correlation coefficient is zero then it means there is no relationship at all between variables. The result of correlation analysis between independent and dependent variable can be presented in Table 4 below:

Table 4. Model summary.

| Model | R | R Square | Adjusted R Square | Std. Error of the Estimate |
|---|---|---|---|---|
| 1 | 0.725[a] | 0.575 | 0.525 | 0.26961 |

Based on Table 4, the strength of the relationship between campus service facility quality variables and user satisfaction is 0.725. If referring to Table 5 below, the relationship or correlation between campus service facility variables and user satisfaction variables is strongly interpreted. In addition to a strong correlation, the pattern of the relationship was positive. This means that if the service quality variables increase, then the satisfaction variable will also rise and vice versa ie if the variable of campus service facilities decreased then the variable of user satisfaction will also decrease [10].

Table 5. Interpretation of Correlation Coefficient.

| Coefficient Interval | Correlation |
|---|---|
| 0.00 – 0.199 | Very Weak |
| 0.20 – 0.399 | Weak |
| 0.40 – 0.599 | Medium |
| 0.60 – 0.799 | Strong |
| 0.80 – 1.000 | Very Strong |

In Table 4, it can be shown that the variation of customer satisfaction variables can be explained by independent variables namely the quality of campus service facilities through the value of coefficient of determination R Square ($R^2$). The result of analysis with statistic obtained value $R^2$ equal to 0.575. This means the variable quality of campus service facilities as free variables contribute 57.5% to the dependent variable that is user satisfaction. Thus there are still 42.5% of the rest can be explained other variables that are not contained in the model. The results of this study reinforce previous research that the positive relationship between service quality and student satisfaction [6] and the influence of service quality on student satisfaction [3][4][5].

## 5. Conclusion
Based on the results of research conducted could be drawn some conclusions are follows:
1. The evaluation of the condition of the quality of the campus service facilities shows a low level of student satisfaction, which can be seen from the gap between the perceptions and expectations of users where for each item there is negative. In other words, the quality of campus service facilities is still far below user expectations.
2. Three of campus service facilities that have the lowest index based on user perception are campus: computer and multimedia laboratory in classroom and wifi network.
3. There is a strong and positive relationship (coefficient R) of 0.725 between the quality of service facilities and student satisfaction.
4. The quality of campus service facilities had an effect on student satisfaction as the user of 57.5%. It means all service facility variables in this study could explain the variation of student satisfaction by 57.5% and the rest about 42.5% would be explained with other variables.
5. Further research suggestions should be examined other variables outside this study which also affect the user satisfaction.

**Acknowledgments**
We would like to thank the research institution that has supported the research activities was being carried out properly.